# Leveraging RNNs and LSTMs for Synchronization Analysis in the Indian Stock Market: A Threshold-Based Classification Approach

Sanjay Sathish

ss219@snu.edu.in

Department of Computer Science and Engineering

School of Engineering

Shiv Nadar Institution of Eminence, Delhi NCR

Dr. Charu C Sharma

charu.sharma@snu.edu.in

Department of Mathematics

School of Natural Sciences

Shiv Nadar Institution of Eminence, Delhi NCR

**Abstract**

Our research presents a new approach for forecasting the synchronization of stock prices using machine learning and non-linear time-series analysis. To capture the complex non-linear relationships between stock prices, we utilize recurrence plots (RP) and cross-recurrence quantification analysis (CRQA). By transforming Cross Recurrence Plot (CRP) data into a time-series format, we enable the use of Recurrent Neural Networks (RNN) and Long Short-Term Memory (LSTM) networks for predicting stock price synchronization through both regression and classification. We apply this methodology to a dataset of 20 highly capitalized stocks from the Indian market over a 21-year period. The findings reveal that our approach can predict stock price synchronization, with an accuracy of 0.98 and F1 score of 0.83 offering valuable insights for developing effective trading strategies and risk management tools.

## 1.Introduction

Understanding the price dynamics with respect to the financial time series has always been a very popular topic of research. A good understanding of the price movements can help in building better predictive models which can be used to construct profitable portfolios. There are many factors that influence the price movement of stocks and one of the key factors is the price change of other stocks (Sharma and Habib, 2019). Over the years, many researchers have developed various models to study the dynamics of the price co-movement of stocks. Among various models, the stock networks based on Pearson's cross-correlation has been widely used to study the dynamics of inter-connectivity of stock prices and their returns (Pan and Sinha, 2007). The only drawback of such models is that they tend to capture only the linear interconnections and may miss out on any non-linear dynamics amongst the price movements of stock (Sharma and Habib, 2019). Mutual information (MI) between two random variables captures the statistical dependence of one variable over the other. Recently researchers have started exploring MI-based stock price networks. Guo, Zhang and Tian(2018) studied Chinese Stock market; Barbi and Prataviera (2019) studied Brazilian equity market; also, in same year Sharma and Habib (2019) studied Indian stock market based on networks constructed using MI. MI between two variables is found useful in capturing non-linear relationships but may not capture complex nonlinear relationships. Also, it does not provide direct insight into the nature of these relationships. This can make it difficult to interpret the results of the model and understand the factors driving stock price predictions.

The autoregressive models for stock price predictions have also been very popular with the research community. Assumption of linearity, stationarity and a fixed probability distribution is a major disadvantage in using these models standalone. With the massive financial data, many researchers have now started exploring machine learning algorithms in stock price prediction and predicting the co-movements of stocks (Sako et al., 2022; Tsang et al., 2018; Hansun and Young, 2021). Deep learning models are used to build non-linear stock prices models with no assumption in terms of linearity, stationarity or probabilistic distributions

Another direction that has been popular amongst the researchers to study the non-linear dynamics of univariate as well as multivariate time series is using recurrence plots (RP) and recurrence quantification analysis (RQA) (Shabani et al, 2023). It is a powerful tool useful for detecting and quantifying nonlinear dynamics in time series data. It can reveal hidden patterns and structures that may not be apparent through traditional linear analysis methods. Eckmann et al. (1987) introduced recurrence plots (RPs) that helped in characterizing dynamical systems without any assumptions. Since then, RQA and RP methods are used in various fields to study the dynamics of the system. Cross recurrence quantification analysis (CRQA) and cross recurrence plots (CRPs) is a generalization of RQA and RP used to study the co-movement of two different time series. These techniques help in detecting phase transitions and critical points in a system, where there is a sudden change in behavior or dynamics. This is useful for studying phenomena such as synchronization, chaos, and criticality (Marwan et.al, 2002). In 2023, Shabani et al proposed a CRP and convolutional neural network (CNN) based method to predict the state of synchronization between two financial stock price time series. They used the CRPs generated over a rolling window as input variables in the CNN formed by two convolutional and one fully connected layer. They tested their method on 66 pairs formed with12 stocks on a 7-year dataset and got satisfactory result in determining the paired movements of stock prices.

Motivated by the earlier works in this paper we present a methodology for predicting the synchronization of stock movements. Our method considers the actual distances between the time points in the phase diagram instead of CRPs and takes these as an input to Recurrent Neural Network (RNN) and Long Short-Term Memory (LSTM) models to predict the synchronization of the paired stock prices. We performed our analysis on 190 pairs obtained from 20 highly capitalized stocks picked from 14 different sectors from Indian Stock market on a period of 21 years.

The paper further is divided into four main sections: data description, methods and methodology, discussion, and conclusion. The data description section provides details about the dataset used in the analysis, while the methods and methodology section give an overview of the methods used. The next section presents the comparative results. Finally, the paper concludes by highlighting the key observations and interpretations from the analysis.

## 2. Data

We worked with the adjusted daily stock price data of 26 highly capitalized stocks from 14 different sectors listed with the National Stock Exchange of India (NSE). The time period considered for the analysis is from January 2003 to December 2023, a total of 5211 days. Entire data was downloaded from Yahoo Finance (https://finance.yahoo.com/). In each business sector, we choose the first two stocks with the highest but comparable capitalization, a strategy supported by financial theory (Shabani et al, 2023). Market sectors naturally group securities and conducting analyses at this level is a common practice to ensure comparability and robustness of results. It is widely recognized that economic variables exhibit asymmetric dynamics across market sectors. However, due to a significant number of consecutive missing data points, we had to remove 6 stocks from our analysis. Table 1 lists the details of the remaining 20 stocks. We consider each of the 190 ($C_2^{20}$) pairs at a time and for each pair we consider first 70% of the data (~3648 data points) as training price data to build the model and next 30% of the data (~1563 data points) as the testing dataset to test the efficiency and accuracy of our proposed model. Each of the $k (k = 1,2, \dots 20)$ training time series was normalized by subtracting their respective means $\mu_k$ and dividing by their respective standard deviations $\sigma_k$. The testing data for each of the respective time series was also normalized but using the parameters $\mu_k$ and $\sigma_k$ obtained from the training dataset.

| Company Symbol | Sector | Market Capitalization (in lakhs) as on 31/03/2023 |
|---|---|---|
| HDFCBANK | FINANCIAL SERVICES | 89808750.012 |
| M&M | AUTOMOBILE | 14408768.56 |
| ULTRACEMCO | CEMENT & CEMENT PRODUCTS | 22003819.65 |
| GRASIM | CEMENT & CEMENT PRODUCTS | 10750221.32 |
| PIDILITIND | CHEMICALS | 11960634.07 |
| LT | CONSTRUCTION | 30416456.17 |
| HINDUNILVR | CONSUMER GOODS | 60157759.88 |
| ITC | CONSUMER GOODS | 47632201.22 |

| RELIANCE | ENERGY | 157706937.8 |
| --- | --- | --- |
| ONGC | ENERGY | 19002511.74 |
| UPL | FERTILISERS & PESTICIDES | 5386735.736 |
| ICICIBANK | FINANCIAL SERVICES | 61248250.45 |
| SIEMENS | INDUSTRIAL MANUFACTURING | 11848655.06 |
| ABB | INDUSTRIAL MANUFACTURING | 7131034.681 |
| TCS | IT | 117305528 |
| INFY | IT | 59239363.15 |
| TATASTEEL | METALS | 12771501.6 |
| SUNPHARMA | PHARMA | 23587862.09 |
| CONCOR | SERVICES | 3535430.454 |
| BHARTIARTL | TELECOM | 41757758.53 |

**Table 1:** List of 20 stocks based on market capitalization considered for analysis

## 3. Methods and Methodology

### 3.1 Recurrence Analysis and Cross Recurrence Quantification Analysis

Recurrence plots (RPs) are a visual tool used to analyze the dynamic behavior of time series data. Introduced by Eckmann et al. in 1987, RPs offer a way to identify and visualize the times at which a dynamical system revisits the same or similar states. This method provides insights into the temporal structure of the data and helps uncover patterns that may not be evident through traditional time series analysis.

A RP of an embedded time series $\{X_i\}_i$ is a two-dimensional, square matrix where both axes represent time indices of the time series. The $(i,j)th$ element of the matrix is defined by:

$$R_{ij}(\epsilon) = \Theta\big(\epsilon - \|X_i - X_j\|\big), \qquad (1)$$

where $\Theta$ is the Heaviside function, $\epsilon$ is a predefined threshold distance, $\|.\|$ is the Euclidean norm. The value of R(i,j) is 1 if the distance between the states at times $t_i$ and $t_j$ is below $\epsilon$, indicating a recurrence, and 0 otherwise. The resulting matrix is binary, with 1s indicating recurrences and 0s indicating non-recurrences. Recurrence quantification analysis (RQA) is a method to quantify the number and duration of recurrences of a dynamical system is occurring in a recurrence plot. Multidimensional RQA(MdRQA) extends the concept of recurrence plot and quantification measures to analyse multiple variables simultaneously (Wallot et al., 2016). For each stock, the two-time series considered for analysis exploring the synchronization are the price and volume series. Figure 1 gives the RP for the *HDFC bank* for Price series (Figure1a), Volume series (Figure 1b) and Price-Volume series (Figure1c).

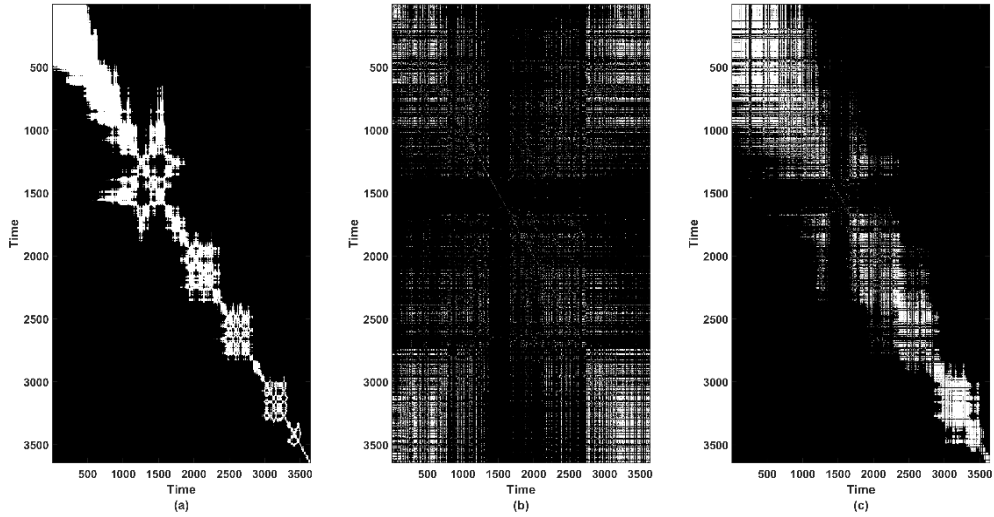

**Fig. 1**. Recurrence plots for HDFC bank

In Fig.1, (a) corresponds to RP for the Price timeseries while (b) corresponds to the Volume timeseries and (c) for the Price-Volume timeseries for the HDFC bank data. Black colour represents 0 and white colour represents 1.

A cross-recurrence plot is a visualization technique, particularly useful for studying the relationships and similarities between two or more time series. (Marwan et.al, 2002) extended the concept of RP to CRP which helped in studying the joint dynamics of a paired time series. A CRP gives a nice visualization of the time dependence of two different time series $\{x_i\}$ and $\{y_j\}$ and captures their interdependence even in the case of very complex datasets. Moreover, CRPs make no assumption of the stationarity of the datasets and therefore they are very useful in the analysis of systems whose dynamics may be changing. The construction of CRP starts by first embedding the four normalized training timeseries $\{x_i^P\}_{i=1}^{3648}$, $\{x_i^V\}_{i=1}^{3648}$, $\{y_j^P\}_{j=1}^{3648}$ and $\{y_j^V\}_{j=1}^{3648}$ in a $D$ dimensional space using a delayed parameter $\tau$, $\{x_i^P\}$ and $\{x_i^V\}$ respectively corresponds to the price and volume series of one stock. Similarly, $\{y_j^P\}$ and $\{y_j^V\}$ respectively corresponds to the price and volume series of the second stock. There are various methods to find an optimal $D$ and $\tau$. For our analysis we fixed $\tau = 1$ and used the False Nearest Neighbors method (Rhodes and Morari, 1997) to find embedding dimension $D_k$ for each time series and then consider a common embedding dimension $D$ as maximum of all 40-time series. Thus, all the 20 timeseries corresponding to price data and 20 timeseries corresponding to volume data were embedded in $R^4$.

After embedding the training time series $\{x_i^P\}_i$ and $\{x_i^V\}_i$ in the embedded phase space, let the respective embedded time series be represented by $\{X_i\}_i$ as

$$X_i = (x_i^P, x_{i+1}^P, x_{i+2}^P, x_{i+3}^P, x_i^V, x_{i+1}^V, x_{i+2}^V, x_{i+3}^V) \quad (2)$$

Next for each of the 190 pairs $\{X_i\}_i$ and $\{Y_j\}_j$ and for a fixed $\epsilon$, we construct the CRPs using the recurrence point matrices $C^{XY}$ (eq 2) and distance matrices $D^{XY}$ (eq 3).

$$C_{i,j}^{XY}(\epsilon) = \Theta(\epsilon - \|X_i - Y_j\|), i,j = 1,2,3\ldots,N \quad (3)$$

$$D_{i,j}^{XY}(\epsilon) = \|X_i - Y_j\|, i,j = 1,2,3\ldots,N \quad (4)$$

where $N$ is the number of time states, $\Theta$ is the Heaviside function i.e.

$$\Theta(x) = 1, if\ x \geq 0\ and\ \Theta(x) = 0, otherwise$$

$\|.\|$ is the Euclidean norm and $\epsilon$ is the threshold distance. Figure 2(a) and 2(b) gives the CRP and distance plots for the HDFC Bank and UPL. We use $D_{i,j}^{XY}$ to predict synchronization between time series $X$ and $Y$ at a given time $t$.

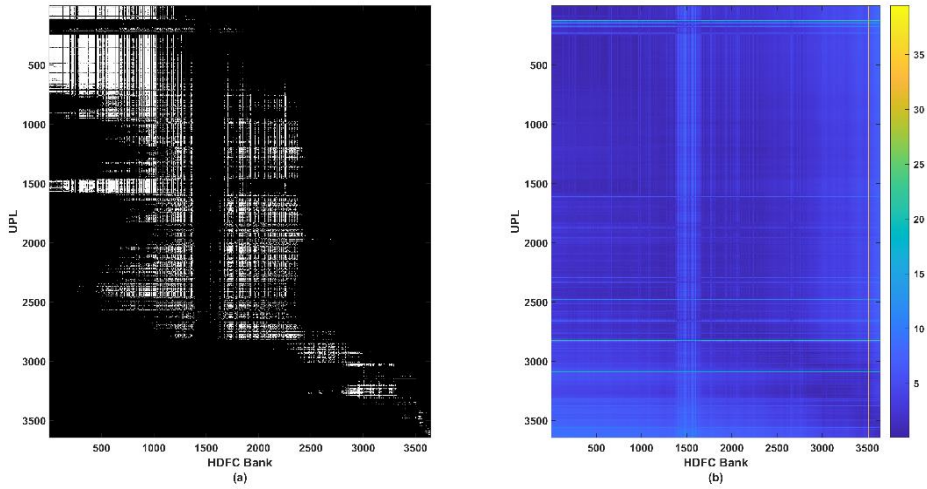

**Fig. 2.** CRP and distance plots for the HDFC Bank and UPL

Fig. 2 (a) corresponds to CRP with black colour representing 0 and white colour representing 1, Fig. 2 (b) is the heatmap corresponding to the distance matrix with smaller distances represented by shades of blue ranging to higher distances in yellow colour.

### 3.2 Time-Series Forecasting

In previous studies, researchers successfully proposed a CRP and convolutional neural network (CNN) based method to predict the state of synchronization between two financial stock price time series (Shabani et al, 2023). They used sub-sampled cross recurrence plots as an input to CNN and predicted the state of synchronization. Motivated by this study we also employed a CNN model on the sub-sample of the CRP matrix for our dataset to predict the synchronization. However, this method underperformed on our dataset with accuracy of

55.47% and F1 score of 0.45. The CNN based method used a single model for all sectors but now we propose to convert the problem into a time-series forecasting, where we try to forecast the distance between 2 stocks.

For each pair of stocks, we converted the distance matrix corresponding to the CRP into a format suitable for regression analysis, we take sub-matrices along the diagonal of size $n \times n$ from the distance matrix and call them as a block. In our experiments, we varied $n$ as 1, 10, 20 and 30.

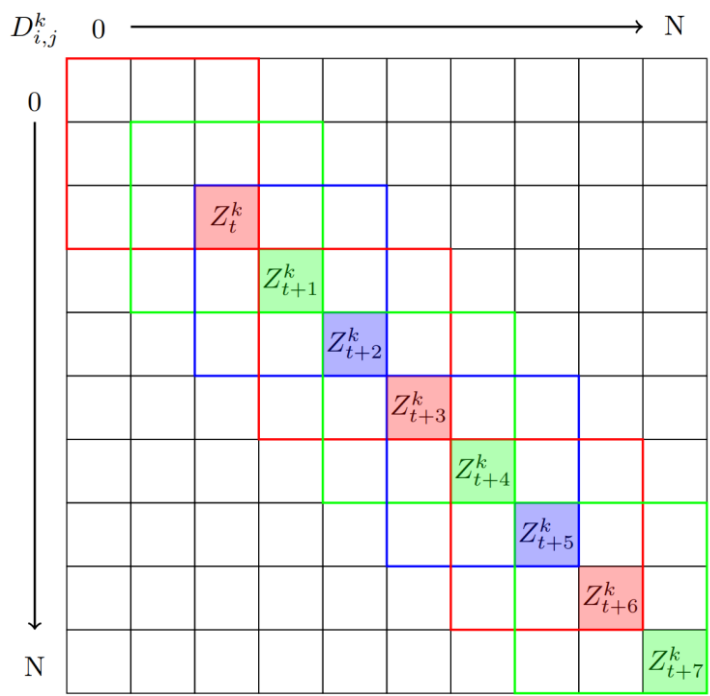

**Fig. 3.** Proposed method in a sample distance CRP diagram.

In the sample distance CRP diagram $D_{i,j}^k$, shown in fig. 3, we take blocks of size $n = 3$. The last cell of each 3x3 block which is highlighted is replaced with $Z_t^k$ by taking the average of the entries in the respective block. The resulting series is $Z^k = (Z_t^k, Z_{t+1}^k, Z_{t+2}^k, .., Z_{t+N-n+1}^k)$, where $k$ denotes the stock pair and varies from 1 to 190.

For each block, we compute the average of the $n^2$ entries and replace the $(n,n)^{th}$ element in the block with this average value. This process transforms the distance matrix of CRP, corresponding to $k^{th}$ pair of stock, into a time series $Z^k$ as shown in figure 3, where each point in the series represents the average distance between the two stocks over $n + 8$ days in an 8-dimensional vector space and $k = 1,2,..190$.

This averaging process reduces the complexity of the distance matrix corresponding to the CRP while preserving the essential information about the recurrence patterns.

## 3.3 Recurrent Neural Networks (RNNs)

RNNs are a class of artificial neural networks specifically designed to handle sequential data. Unlike traditional feedforward neural networks, which process information layer-by-layer in a one-way direction, RNNs incorporate loops within their architecture. These loops allow RNNs to process information from previous steps along with the current Input. This internal memory enables RNNs to excel in tasks involving temporal dependencies, such as time series forecasting, where the output depends on the current input along with the sequence of previous inputs. RNNs can be used to forecast time series data (Graves et al., 2013), such as health conditions (Lipton et al., 2015), weather patterns (Li et al., 2019; Ren et al., 2021; Chattopadhyay et al., 2020), energy consumption (Bedi et al., 2020; Kahn et al., 2020) and more.

RNNs have been widely used in financial markets (Sako et al., 2022; Tsang et al., 2018). Financial time series forecasting aims to predict future stock prices (Hansun and Young, 2021) or exchange rates (Shen et al.,2020). RNNs are particularly well-suited for this task because they can consider historical data points, such as past closing prices, trading volumes, and economic indicators, to make informed predictions about future values. RNNs can be applied not only to forecasting but also to other tasks such as algorithmic trading (Lei et al., 2020), where the model can be trained to automatically generate trading signals based on historical price data and other relevant information. Furthermore, RNNs can be used for risk assessment (Chen and Shen, 2020), by analyzing time series data of financial instruments to evaluate potential risks associated with investments.

For our study, we employ a simple RNN model using 4 layers. We use 3 simple RNN layers and one dense layer. The optimal number of layers was determined by experimenting with different layer counts and evaluating their performance on our dataset. The input to the RNN model is the time series $Z^k$, as explained in section 2.3, corresponding to the average block distance rolled over by a day. The RNN model processes the transformed time series $Z^k$, where each point represents the average block distance in the distance matrix. The model aims to predict these values based on historical patterns in the data. The output would be a sequence of predicted average distances for the time series pairs.

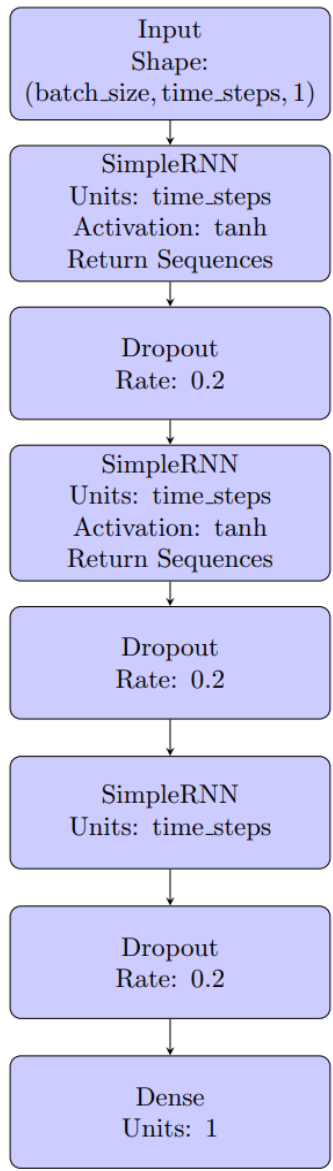

**Fig. 4.** Architecture diagram of RNN model used. This model comprises multiple SimpleRNN layers, each followed by dropout layers for regularization followed by a dense output layer to generate final predictions.

### 3.4 Long Short-Term Memory (LSTM) Networks

LSTMs address the vanishing gradient problem by introducing a gating mechanism within their architecture. LSTMs are a specific type of RNN unit designed to learn long-term dependencies. They achieve this using three primary gates:

Forget Gate: This gate determines which information from the previous cell state (short-term memory) should be forgotten. It analyzes the current input and the previous hidden state to decide what past information remains relevant.

Input Gate: This gate controls the flow of new information from the current input into the cell state. It determines what new information needs to be stored in the cell's memory based on the current input and the previous hidden state.

Output Gate: This gate regulates the flow of information from the current cell state to the output layer. It decides what information from the cell state should be used to influence the current output and future predictions.

By remembering and forgetting information through these gates, LSTMs can learn long-term dependencies within sequential data. This makes them particularly well-suited for tasks where capturing long-range relationships is crucial, such as financial time series forecasting over extended periods. After the LSTM processes the input sequence, the final output layer produces a prediction for the next value in the time series, which can then be compared to actual values for evaluation. In summary, the output of this LSTM model for each input sequence segment $Z^k$ is a set of predicted values for the next point in the sequence.

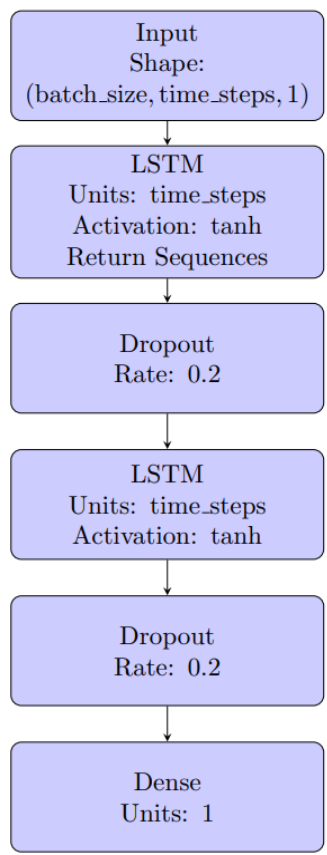

**Fig. 5.** Architecture diagram of LSTM model used. This model comprises two LSTM layers, each followed by dropout layers for regularization followed by a dense output layer to produce the final predictions.

### 3.5 Choice of Hyperparameters

We use common hyperparameters for both the RNN and LSTM models used in our study. For the activation function, we use the tanh activation function. It introduces non-linearity into the network and maps input values to a range between -1 and 1, aiding in training stability and enabling the network to model complex relationships in the data. For the learning rate optimizer algorithm, we use the adam optimizer which dynamically adjusts individual parameter learning rates during training based on historical gradients, accelerating convergence and reduce the risk of getting stuck in local minima. Finally, for our loss function, we use the mean squared error (MSE). MSE is a widely used regression loss function (Hansun and Young, 2021; Sako et al., 2022). It calculates the average squared difference between the model's predictions and the ground truth labels. Minimizing MSE during training encourages the model to produce outputs that closely match the actual target values.

### 3.6 Evaluating Time-Series Forecasting

We employ RNNs and LSTM networks to model the transformed time series data. As explained in section 2.4 and 2.5, both RNNs and LSTMs are well-suited for time series analysis due to their ability to capture temporal dependencies.

We perform experiments varying the block size $n$ (1, 10, 30, 50) and the number of time steps (20, 50) considered by the RNN and LSTM models. For each experiment, we train the models using the training set for each of the 190 pairs of stocks and evaluate their performance on the test set. The evaluation metrics used are:

- **R-squared ($R^2$):** Measures the proportion of the variance in the dependent variable that is predictable from the independent variables.

$$R^2 = 1 - \frac{\sum_i (y_i - \hat{y}_i)^2}{\sum_i (y_i - \bar{y})^2}$$

$$y_i = Actual\ value$$
$$\hat{y}_i = Predicted\ value$$
$$\bar{y} = Mean\ of\ all\ actual\ values$$

- **Mean Absolute Percentage Error (MAPE):** Measures the accuracy of the model as a percentage.

$$MAPE = \frac{1}{n} \sum_i^n \left|\frac{y_i - \hat{y}_i}{y_i}\right|$$

$$y_i = Actual\ value$$

$$\hat{y}_i = Predicted\ value$$
$$n = Length\ of\ dataset$$

- **Mean Absolute Error (MAE):** Measures the average magnitude of the errors in a set of predictions

$$MAE = \frac{1}{n}\sum_i^n |y_i - \hat{y}_i|$$

$$y_i = Actual\ value$$
$$\hat{y}_i = Predicted\ value$$
$$n = Length\ of\ dataset$$

- **Root Mean Squared Error (RMSE):** Measures the square root of the average of squared differences between prediction and actual observation.

$$RMSE = \sqrt{\sum_i^n \frac{(y_i - \hat{y}_i)^2}{n}}$$

$$y_i = Actual\ value$$
$$\hat{y}_i = Predicted\ value$$
$$n = Length\ of\ dataset$$

These metrics have been widely used in evaluating regression models and time-series forecasting problems (Hansun and Young, 2021; Sako et al., 2022; Esparza-Gómez et al., 2023).

### 3.7 Classification

With the optimal settings determined from evaluating the results of the time-series forecasting experiments, we continue into the classification task. We define thresholds corresponding to recurrence rates of 20%, 25%, 30%, 35%, 40% and 45% for each stock pair, to classify the distances into binary labels using the Heaviside function. Specifically, we convert the continuous distance measures into binary classes, where a distance below the threshold is classified as 1 (synchronous) and above the threshold as 0 (non-synchronous). Similar approaches of using RNNs and LSTMs for classification have been used in various domains, including speech recognition, sentiment analysis, and medical diagnosis, where RNNs and LSTMs have demonstrated their efficacy in handling complex, sequential data for classification tasks (Lipton et al., 2015; Graves et al., 2013). Moreover, the use of thresholds for classification has been explored in the context of anomaly detection, where the goal is to identify deviations from normal behavior. By setting appropriate thresholds, RNNs and LSTMs can classify these anomalies accurately, making them useful for applications in fraud detection, network security, and system monitoring (Malhotra et al., 2015; Chauhan & Vig, 2015). The evaluation metrics used are:

- **Accuracy:** The ratio of correctly predicted observation to the total observations.

$$Accuracy = \frac{TP + TN}{TP + FP + FN + TN}$$

where TP is True Positives, TN is True Negatives, FP is False Positives and FN is False Negatives.

- **F1 Score:** The harmonic mean of precision and recall, providing a single metric that balances both concerns.

$$F1\ score = \frac{2 \times Precision \times Recall}{Precision + Recall}$$

- **Precision:** The ratio of correctly predicted positive observations to the total predicted positives.

$$Precision = \frac{TP}{TP + FP}$$

- **Recall:** The ratio of correctly predicted positive observations to all observations in the actual class.

$$Recall = \frac{TP}{TP + FN}$$

Since the class distribution can be imbalanced, especially with the minority class (class label 1), we note the percentage of true class 1 labels in the experiments' dataset.

## 4. Results and Discussion

### 4.1 Evaluating Time-Series Forecasting

We compute the average and standard deviation of each metric (section 3.6) across all 190 stock pairs for each experiment to identify the best-performing model configuration in table 1 and table 2 respectively. It can be inferred from table 1 that the LSTM model generally perform better than RNNs, particularly with larger block sizes of 20-30, as evidenced by slightly higher R² values and lower error metrics. There is some variability in performance between the time steps of 50 and 20, with longer time steps often providing better historical context, resulting in higher accuracy, particularly with larger block sizes. Additionally, we can see that larger block sizes lead to improved performance for both models. As the block size increases, average R² values increase, and error metrics (MAPE, RMSE, MAE) decrease, suggesting that larger blocks capture more significant patterns in the data. From table 3, it can be inferred that smaller block sizes (example: block size of 1) show higher standard deviation in metrics, indicating more variability in model performance. As the block size increases, the standard deviations generally decrease, suggesting more consistent predictions with larger block sizes. However, this trend is not uniform, as seen with the RNN at a block size of 30, where the standard deviation increases for RMSE and MAE from time step of 20 to time step of 50. The RNN models generally exhibit higher variability (as indicated by higher standard deviations) compared to LSTM models, especially for block sizes of 1 and 10. However, at block sizes of 20 and 30, the LSTM model show higher variability compared to the RNN model.

| Block | Time Steps | Model | R² | MAPE | RMSE | MAE |
|---|---|---|---|---|---|---|
| 1 | 50 | rnn | 0.79 | 0.2 | 1.67 | 1.02 |
| 1 | 20 | rnn | 0.79 | 0.2 | 1.69 | 1.04 |
| 10 | 50 | rnn | 0.97 | 0.07 | 0.51 | 0.37 |
| 10 | 20 | rnn | 0.96 | 0.07 | 0.6 | 0.42 |
| 20 | 50 | rnn | 0.98 | 0.05 | 0.35 | 0.27 |
| 20 | 20 | rnn | 0.98 | 0.05 | 0.41 | 0.31 |
| 30 | 50 | rnn | 0.98 | 0.04 | 0.32 | 0.25 |
| 30 | 20 | rnn | 0.98 | 0.05 | 0.38 | 0.28 |
| 1 | 50 | lstm | 0.79 | 0.18 | 1.68 | 0.99 |
| 1 | 20 | lstm | 0.79 | 0.18 | 1.67 | 0.97 |
| 10 | 50 | lstm | 0.98 | 0.05 | 0.46 | 0.31 |
| 10 | 20 | lstm | 0.97 | 0.07 | 0.54 | 0.37 |
| 20 | 50 | lstm | 0.98 | 0.04 | 0.33 | 0.24 |
| 20 | 20 | lstm | 0.98 | 0.05 | 0.41 | 0.29 |
| 30 | 50 | lstm | 0.98 | 0.04 | 0.32 | 0.24 |
| 30 | 20 | lstm | 0.98 | 0.05 | 0.37 | 0.27 |

**Table 2**. Performance metrics for RNN and LSTM models across various time steps and window sizes. The table includes the average R², MAPE, RMSE, and MAE across the 190 stock pairs taken

| Block | Time Steps | Model | R² | MAPE | RMSE | MAE |
|---|---|---|---|---|---|---|
| 1 | 50 | rnn | 0.15 | 0.15 | 0.89 | 0.73 |
| 1 | 20 | rnn | 0.13 | 0.15 | 0.83 | 0.65 |
| 10 | 50 | rnn | 0.04 | 0.04 | 0.29 | 0.25 |
| 10 | 20 | rnn | 0.03 | 0.05 | 0.33 | 0.28 |
| 20 | 50 | rnn | 0.02 | 0.03 | 0.22 | 0.19 |
| 20 | 20 | rnn | 0.03 | 0.03 | 0.26 | 0.22 |
| 30 | 50 | rnn | 0.03 | 0.03 | 0.21 | 0.18 |
| 30 | 20 | rnn | 0.03 | 0.03 | 0.27 | 0.22 |
| 1 | 50 | lstm | 0.13 | 0.12 | 0.81 | 0.61 |
| 1 | 20 | lstm | 0.1 | 0.11 | 0.76 | 0.52 |
| 10 | 50 | lstm | 0.03 | 0.03 | 0.25 | 0.2 |
| 10 | 20 | lstm | 0.05 | 0.04 | 0.3 | 0.25 |
| 20 | 50 | lstm | 0.04 | 0.03 | 0.26 | 0.21 |
| 20 | 20 | lstm | 0.02 | 0.02 | 0.21 | 0.17 |
| 30 | 50 | lstm | 0.04 | 0.03 | 0.32 | 0.27 |
| 30 | 20 | lstm | 0.02 | 0.02 | 0.23 | 0.19 |

**Table 3.** Performance metrics for RNN and LSTM models across various time steps and window sizes. The table includes the standard deviation of R², MAPE, RMSE, and MAE across the 190 stock pairs taken

| Block | Time Steps | Model | R² | MAPE | MAE | RMSE |
|---|---|---|---|---|---|---|
| 1 | 50 | rnn | 0.19 | 0.76 | 0.71 | 0.53 |
| 1 | 20 | rnn | 0.16 | 0.75 | 0.62 | 0.49 |
| 10 | 50 | rnn | 0.04 | 0.63 | 0.68 | 0.57 |
| 10 | 20 | rnn | 0.03 | 0.65 | 0.67 | 0.55 |
| 20 | 50 | rnn | 0.03 | 0.66 | 0.72 | 0.63 |
| 20 | 20 | rnn | 0.03 | 0.58 | 0.72 | 0.63 |
| 30 | 50 | rnn | 0.03 | 0.59 | 0.73 | 0.66 |
| 30 | 20 | rnn | 0.03 | 0.59 | 0.79 | 0.72 |
| 1 | 50 | lstm | 0.17 | 0.65 | 0.62 | 0.48 |
| 1 | 20 | lstm | 0.13 | 0.61 | 0.54 | 0.46 |
| 10 | 50 | lstm | 0.03 | 0.57 | 0.64 | 0.54 |
| 10 | 20 | lstm | 0.05 | 0.61 | 0.68 | 0.55 |
| 20 | 50 | lstm | 0.04 | 0.71 | 0.88 | 0.78 |
| 20 | 20 | lstm | 0.02 | 0.48 | 0.58 | 0.5 |
| 30 | 50 | lstm | 0.04 | 0.79 | 1.15 | 1.02 |
| 30 | 20 | lstm | 0.02 | 0.51 | 0.69 | 0.62 |

**Table 4**. Performance metrics for RNN and LSTM models across various time steps and window sizes. The table includes the CV of R², MAPE, RMSE, and MAE across the 190 stock pairs taken

To gain a deeper understanding of the model performance for different prediction accuracy levels, we examined the distribution of MAPE scores for both models. Established reference ranges (Montaño Moreno et al., 2013; Hansun & Young, 2021) were used to categorize prediction accuracy, as shown in table 5.

| Block | Time Steps | Model | Highly Accurate Forecasting | Good Forecasting | Reasonable Forecasting | Inaccurate Forecasting |
|---|---|---|---|---|---|---|
| 1 | 50 | rnn | 17 | 127 | 39 | 7 |
| 1 | 20 | rnn | 14 | 120 | 50 | 6 |
| 10 | 50 | rnn | 162 | 25 | 3 | 0 |
| 10 | 20 | rnn | 157 | 27 | 6 | 0 |
| 20 | 50 | rnn | 177 | 13 | 0 | 0 |
| 20 | 20 | rnn | 178 | 11 | 1 | 0 |
| 30 | 50 | rnn | 185 | 5 | 0 | 0 |
| 30 | 20 | rnn | 181 | 9 | 0 | 0 |
| 1 | 50 | lstm | 17 | 129 | 39 | 5 |
| 1 | 20 | lstm | 14 | 134 | 39 | 3 |
| 10 | 50 | lstm | 173 | 16 | 1 | 0 |
| 10 | 20 | lstm | 169 | 17 | 4 | 0 |
| 20 | 50 | lstm | 183 | 5 | 2 | 0 |
| 20 | 20 | lstm | 182 | 8 | 0 | 0 |
| 30 | 50 | lstm | 185 | 4 | 1 | 0 |
| **30** | **20** | **lstm** | **182** | **8** | **0** | **0** |

**Table 5**. Results for RNN and LSTM models across different blocks and window sizes. The table categorizes the forecast accuracy into four levels: Highly Accurate Forecasting, Good Forecasting, Reasonable Forecasting, and Inaccurate Forecasting, showing the count of predictions in each category for various model configurations.

The coefficient of variation (CV) is then calculated for each metric across the datasets for each model. CV is defined as the ratio of the standard deviation to the mean, providing a standardized measure of dispersion relative to the mean. The formula for CV is given by:

$$Coefficient\ of\ Variance\ (CV) = \frac{Standard\ Deviation\ (\sigma)}{Mean\ (\mu)}$$

By using CV, we can compare the variability of different metrics regardless of their scales. This is particularly useful in ensuring that metrics with different units or scales can be compared on a common basis. Once the CVs for all metrics are computed, we rank the models based on their CV values for each metric. Lower CV values indicate more consistent performance across datasets, so models with lower CVs are ranked higher. This ranking process allows us to identify which models demonstrate stable performance and which ones show greater variability. Finally, to determine an overall ranking for each model, we aggregate the ranks across all metrics. The overall rank is calculated by taking the average of the individual ranks for each metric. This aggregated ranking provides a comprehensive measure of each model's performance consistency across all metrics, facilitating a more holistic comparison of model performance.

| Block | Time Steps | Model | R² Score CV | MAPE CV | MAE CV | RMSE CV | Average rank |
|---|---|---|---|---|---|---|---|
| **20** | **20** | **lstm** | **0.02** | **0.48** | **0.58** | **0.5** | **2.25** |
| 1 | 20 | lstm | 0.13 | 0.61 | 0.54 | 0.46 | 5.5 |
| 10 | 50 | lstm | 0.03 | 0.57 | 0.64 | 0.54 | 5.5 |
| 30 | 20 | lstm | 0.02 | 0.51 | 0.69 | 0.62 | 5.5 |
| 10 | 20 | rnn | 0.03 | 0.65 | 0.67 | 0.55 | 7.75 |
| 20 | 20 | rnn | 0.03 | 0.58 | 0.72 | 0.63 | 8 |
| 1 | 50 | lstm | 0.17 | 0.65 | 0.62 | 0.48 | 8 |
| 1 | 20 | rnn | 0.16 | 0.75 | 0.62 | 0.49 | 8.5 |
| 10 | 20 | lstm | 0.05 | 0.61 | 0.68 | 0.55 | 8.5 |
| 30 | 50 | rnn | 0.03 | 0.59 | 0.73 | 0.66 | 9 |
| 10 | 50 | rnn | 0.04 | 0.63 | 0.68 | 0.57 | 9.25 |
| 20 | 50 | rnn | 0.03 | 0.66 | 0.72 | 0.63 | 9.5 |
| 30 | 20 | rnn | 0.03 | 0.59 | 0.79 | 0.72 | 9.75 |
| 1 | 50 | rnn | 0.19 | 0.76 | 0.71 | 0.53 | 11.5 |
| 20 | 50 | lstm | 0.04 | 0.71 | 0.88 | 0.78 | 13 |
| 30 | 50 | lstm | 0.04 | 0.79 | 1.15 | 1.02 | 14.5 |

**Table 6.** Ranking of RNN and LSTM models based on performance metrics using CV, across different blocks and window sizes. The table includes R² Score CV, MAPE CV, MAE CV, RMSE CV, and the Average Rank.

From table 6, we can see that by taking block size of 20 and window size of 20 and using the LSTM model, we achieve the highest average rank.

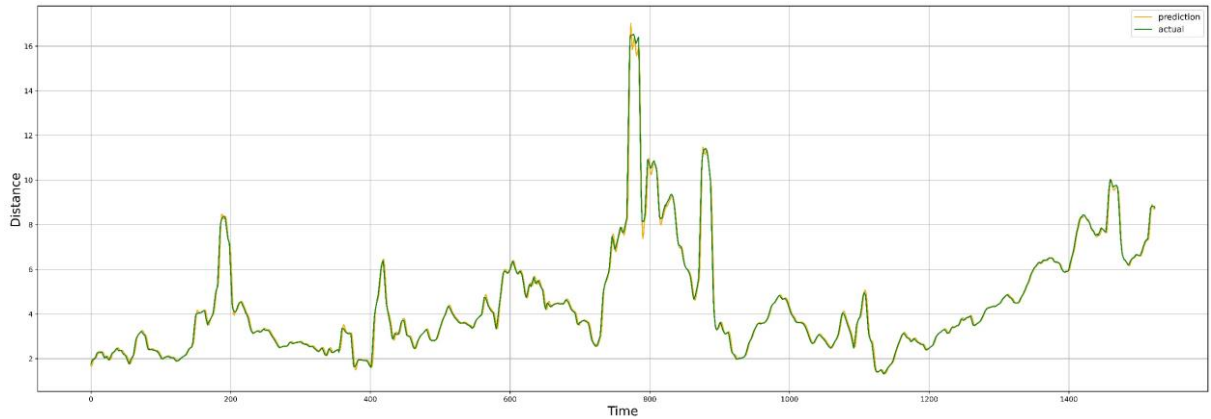

**Fig. 6.** Comparison of actual and predicted distance for HDFCBANK and UPL taking window size of 20 and block size of 20 using the LSTM model. The predicted distances are in yellow, and the actual distances are in green.

### 4.2 Classification Evaluation

From table 7, we can see that the relatively low standard deviations (σ) in accuracy and precision across recurrence rates indicate that the model's performance is consistent and not highly sensitive to changes in these parameters. However, the higher standard deviations in recall and F1 scores, particularly at lower recurrence rates as observed in table 7 (a)-(d), highlight variability in the model's ability to detect positive cases. This variability decreases as the recurrence rate increases, suggesting that a higher recurrence rate provides a more stable performance.

|  | μ | σ | Min | Max | Median |
|---|---|---|---|---|---|
| accuracy | 0.99 | 0.02 | 0.86 | 1 | 0.99 |
| precision | 0.91 | 0.16 | 0 | 1 | 0.97 |
| recall | 0.7 | 0.32 | 0 | 1 | 0.84 |
| f1 | 0.74 | 0.29 | 0 | 1 | 0.88 |
| %class 1 | 0.05 | 0.06 | 0 | 0.27 | 0.02 |

(a) Recurrence rate = 20%

|  | μ | σ | Min | Max | Median |
|---|---|---|---|---|---|
| accuracy | 0.98 | 0.02 | 0.86 | 1 | 0.99 |
| precision | 0.94 | 0.09 | 0.5 | 1 | 0.97 |
| recall | 0.71 | 0.31 | 0 | 1 | 0.85 |
| f1 | 0.75 | 0.29 | 0 | 1 | 0.89 |
| %class 1 | 0.07 | 0.08 | 0 | 0.33 | 0.05 |

(b) Recurrence rate = 25%

|  | μ | σ | Min | Max | Median |
|---|---|---|---|---|---|
| accuracy | 0.98 | 0.02 | 0.88 | 1 | 0.99 |

|  | | | Min | Max | Median |
|---|---|---|---|---|---|
| precision | 0.94 | 0.1 | 0.19 | 1 | 0.97 |
| recall | 0.79 | 0.24 | 0 | 1 | 0.9 |
| f1 | 0.83 | 0.2 | 0 | 1 | 0.91 |
| %class 1 | 0.1 | 0.1 | 0 | 0.41 | 0.06 |

(c) Recurrence rate = 30%

|  | $\mu$ | $\sigma$ | Min | Max | Median |
|---|---|---|---|---|---|
| accuracy | 0.98 | 0.02 | 0.88 | 1 | 0.99 |
| precision | 0.94 | 0.1 | 0.19 | 1 | 0.97 |
| recall | 0.79 | 0.24 | 0 | 1 | 0.9 |
| f1 | 0.83 | 0.2 | 0 | 1 | 0.91 |
| %class 1 | 0.1 | 0.1 | 0 | 0.41 | 0.06 |

(d) Recurrence rate = 35%

|  | $\mu$ | $\sigma$ | Min | Max | Median |
|---|---|---|---|---|---|
| accuracy | 0.98 | 0.02 | 0.9 | 1 | 0.99 |
| precision | 0.94 | 0.12 | 0 | 1 | 0.98 |
| recall | 0.83 | 0.25 | 0 | 1 | 0.94 |
| f1 | 0.86 | 0.22 | 0 | 1 | 0.94 |
| %class 1 | 0.16 | 0.14 | 0 | 0.57 | 0.13 |

(e) Recurrence rate = 40%

|  | $\mu$ | $\sigma$ | Min | Max | Median |
|---|---|---|---|---|---|
| accuracy | 0.98 | 0.02 | 0.88 | 1 | 0.99 |
| precision | 0.94 | 0.1 | 0.19 | 1 | 0.97 |
| recall | 0.79 | 0.24 | 0 | 1 | 0.9 |
| f1 | 0.83 | 0.2 | 0 | 1 | 0.91 |
| %class 1 | 0.1 | 0.1 | 0 | 0.41 | 0.06 |

(f) Recurrence rate = 45%

**Table 7.** Classification results for LSTM model with block size 20 and window size of 20 across various recurrence rates. This table presents the mean ($\mu$), standard deviation ($\sigma$), minimum, maximum, and median values for accuracy, precision, recall, F1 scores, and support metrics across the 190 pairs of stocks taken.

| Recurrence rate | Accuracy CV | Precision CV | Recall CV | F1 CV | %class 1 CV | Average Rank |
|---|---|---|---|---|---|---|
| **45%** | **0.02** | **0.09** | **0.26** | **0.23** | **0.85** | **1.4** |
| 20% | 0.02 | 0.13 | 0.31 | 0.25 | 0.9 | 3 |
| 25% | 0.02 | 0.1 | 0.3 | 0.24 | 1.05 | 3.4 |
| 30% | 0.02 | 0.13 | 0.34 | 0.28 | 0.96 | 4 |

| | | | | | |
|---|---|---|---|---|---|
| 40% | 0.02 | 0.1 | 0.44 | 0.39 | 1.15 | 4.2 |
| 35% | 0.02 | 0.18 | 0.46 | 0.39 | 1.29 | 5 |

**Table 8.** Ranking of Recurrence Rates Based on Coefficient of Variation Across Performance Metrics.

In our analysis to determine the optimal recurrence rate for the LSTM model, we evaluated the performance of different recurrence rates based on several metrics: accuracy, precision, recall, F1 score, and their associated CV. Like the process followed in section 4.1, we first calculated the CV for each metric and then ranked them for all recurrence rates, with lower CV values receiving higher ranks. We computed the average rank across all metrics for each recurrence rate to obtain a consolidated performance measure in table 8. Upon ranking the recurrence rates based on their average ranks, we found that a recurrence rate of 45% achieved the highest rank. This indicates that, on average, the 45% recurrence rate provided the most consistent and favorable performance across the evaluated metrics, making it the best-performing choice for our LSTM model.

**4.3 Comparing Distance Graph with Normalized Price Graphs**

Figure 7 presents two subplots for the stock pair "HDFCBANK_UPL". Subplot (a) illustrates the comparison between the predicted and actual distances over time, with shaded regions indicating specific time intervals. Subplot (b) displays the normalized price trends of HDFCBANK and UPL from January 2003 to December 2023, highlighting different periods with grey, red, and green shaded regions.

Thus, by converting CRP data into a time-series format and employing Recurrent Neural Networks (RNN) and Long Short-Term Memory (LSTM) networks, we were able to capture complex dynamics between stock prices. The application of our methodology to a dataset of 20 top stocks from the Indian market over 21 years demonstrated its superior performance in forecasting and classifying stock price co-movements. This research offers valuable insights for enhancing financial modeling and portfolio optimization.

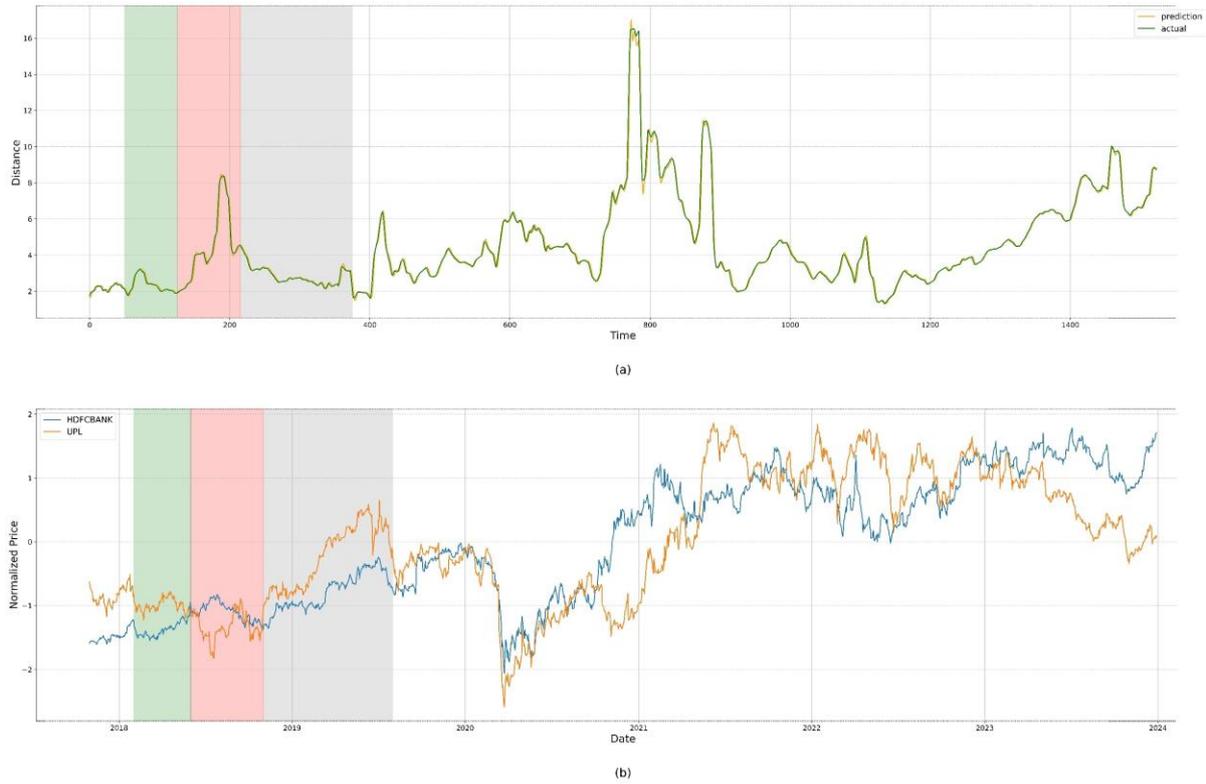

**Fig. 7.** Comparison of Predicted and Actual Stock Distances with Normalized Price Trends.

## 5. Conclusion

Our work proposes a new method of predicting the co-movement of two multidimensional time series with respect to the stock movement. This method is based upon the deep learning models (LSTM and RNN) applied on the cross-recurrence quantification plots. We converted the CRP data into a time-series format and employed RNN and LSTM networks to capture complex dynamics between stock prices. The model is tested on the daily prices and volume data for 20 highly capitalized stocks from 14 different sectors listed with NSE from the year 2003 to 2023. For our study, we employed a 4-layered RNN model and a 3 layered LSTM model. Based upon our wide analysis, we showed that LSTM based deep learning technique, with an accuracy of 98% and F1 score of 0.83, is successfully able to predict the synchronization of two time series on a daily basis. The correct prediction of stock movement can help in enhancing financial modelling and building good trading strategies.

## 6. Data Availability Statement

The datasets and codes used in this study are openly available on GitHub at the following link: https://github.com/SanjaySathish/Cross-Recurrence-Quantification-Analysis.git. This repository contains all the necessary data and scripts required to reproduce the results and analyses presented in this paper.